\documentclass[lettersize,journal]{IEEEtran}
\usepackage{amsmath,amsfonts,amssymb}
\usepackage{algorithmic}
\usepackage{algorithm}
\usepackage{array}
\usepackage[caption=false,font=normalsize,labelfont=sf,textfont=sf]{subfig}
\usepackage{textcomp}
\usepackage{stfloats}
\usepackage{url}
\usepackage{verbatim}
\usepackage{graphicx}
\usepackage{cite}
\usepackage{mathtools}
\usepackage{dsfont}
\DeclareMathOperator*{\argmax}{arg\,max}

\begin{document}

\title{Photon Counting Histogram Expectation Maximization Algorithm for Characterization of Deep Sub-Electron Read Noise Sensors}

\author{
Aaron Hendrickson and David P. Haefner
\thanks{A. Hendrickson is with the U.S.~Navy, NAWCAD, Maryland, U.S.A.}
\thanks{D. Haefner is with the U.S.~Army, C51SR Center, Virginia, U.S.A.}
\thanks{Manuscript received Month DD, 20XX; revised Month DD, 20XX.}
} 

\markboth{Journal of the Electron Devices Society,~Vol.~X, No.~X, Month~20XX}%
{}

\IEEEpubid{0000--0000/00\$00.00~\copyright~20XX IEEE}

\maketitle

\begin{abstract}
We develop a novel algorithm for characterizing Deep Sub-Electron Read Noise (DSERN) image sensors. This algorithm is able to simultaneously compute maximum likelihood estimates of quanta exposure, conversion gain, bias, and read noise of DSERN pixels from a single sample of data with less uncertainty than the traditional photon transfer method. Methods for estimating the starting point of the algorithm are also provided to allow for automated analysis. Demonstration through Monte Carlo numerical experiments are carried out to show the effectiveness of the proposed technique. In support of the reproducible research effort, all of the simulation and analysis tools developed are available on the MathWorks file exchange \cite{PCHEM_code}.
\end{abstract}

\begin{IEEEkeywords}
clustering algorithms, conversion gain, DSERN, expectation maximization, Gaussian mixture, PCH, PCH-EM, photon counting, photon transfer, quanta exposure, read noise.
\end{IEEEkeywords}


\section{Introduction}
\IEEEPARstart{A}{dvances} in image sensor technology have resulted in pixels with sufficiently low read noise; enabling the ability to discern the presence of individual electrons without the need for avalanche gain or electron multiplication \cite{fossum_2013,fossum_2015,fossum_2015_2,fossum_2016}. Sensors with this property, aptly named Deep Sub-Electron Read Noise (DSERN) sensors, open the door to new applications for the CMOS sensor architecture in ultra low-light imaging environments. With the first DSERN sensors now commercially available, methodologies for characterizing these devices has become an emerging topic of interest \cite{starkey_2016,Nakamoto_2022}.

To date, three methods for characterizing DSERN sensors have been developed including the Photon Transfer (PT) method \cite{beecken_96,janesick_2007,hendrickson_22}, Photon Counting Histogram (PCH) method \cite{starkey_2016}, and a third method based on Maximum Likelihood Estimation (MLE) \cite{Nakamoto_2022}. The work by Fossum \& Starkey \cite{starkey_2016}--and subsequently Nakamoto \& Hotaka \cite{Nakamoto_2022}--show promise that the PCH and MLE methods out-perform the traditional PT method in accuracy and precision when applied to DSERN pixels.

While promising, both methods present challenges related to numerical stability, computational cost, and/or autonomy. In this correspondence we present a new method in the form of the PCH Expectation Maximization (PCH-EM) algorithm, which incorporates attributes of both the PCH and MLE methods to enable a fully automated characterization technique for maximum likelihood estimation of quanta exposure, conversion gain, bias, and read noise of DSERN pixels.

Prior to presenting the PCH-EM algorithm we must, however, first develop a statistical model for the digital signal produced by DSERN pixels.



\section{The Photon Counting Distribution Model}
\label{sec:PCD_model}

We begin by deriving the Photon Counting Distribution (PCD) as a model of the probability density for data generated by DSERN pixels. The number of free-electrons generated in a DSERN pixel when exposed to a constant rate of impinging photons  can be modeled by the Poisson random variable $K\sim\operatorname{Poisson}(H)$. Here, $H=H_\gamma+H_d$ denotes the \emph{quanta exposure} describing the expected number of free-electrons generated in the pixel per integration time, which is further decomposed into $H_\gamma$ (the expected number of photoelectrons generated by interacting photons) and $H_d$ (the expected number of free-electrons generated by thermal contributions, i.e.~dark current). The act of sensing the electron signal introduces a continuous read noise component $R\sim\mathcal N(0,\sigma_R^2)$, where $\sigma_R$ is the input-referred analog read noise in $(e\text{-})$. The pixel output signal in Digital Numbers $(\mathrm{DN})$ can thus be represented by the random variable
\begin{equation}
    \label{eq:PCD_RV_definition}
    X=\lceil (K+R)/g+\mu\rfloor,
\end{equation}
where $g$ is the conversion gain in $(e\text{-}/\mathrm{DN})$, $\mu$ is the bias (DC offset) in units of $(\mathrm{DN})$, and $\lceil\cdot\rfloor$ denotes rounding to the nearest integer.

To derive the PCD, we will model the act of quantization (rounding) as a simple additive noise process so that $X|K=k\sim\mathcal N(\mu+k/g,\sigma^2)$, where $\sigma=(\sigma_R^2/g^2+\sigma_Q^2)^{1/2}$ is the combined read and quantization noise in $(\mathrm{DN})$. To avoid confusion, we note that when characterizing an image sensor, the quantity $\sigma$ is commonly referred to as the read noise with $\sigma g$ being its corresponding value in electron units. We may now obtain the PCD by integrating the join density $f_{XK}(x,k)=\mathsf P(K=k)f_{X|K}(x|k)$ w.r.t. $k$ yielding
\begin{equation}
    \label{eq:X_density}
    f_X(x)=\sum_{k=0}^\infty\frac{e^{-H}H^k}{k!}\phi(x;\mu+k/g,\sigma^2),
\end{equation}
where $\phi(x;\mu,\sigma^2)=\frac{1}{\sqrt{2\pi}\sigma}\exp(-(x-\mu)^2/2\sigma^2)$ is the Gaussian probability density with mean $\mu$ and variance $\sigma^2$. For notational purposes we will use $X\sim\operatorname{PCD}(H,g,\mu,\sigma^2)$ to denote a PCD random variable with parameter $\theta=(H,g,\mu,\sigma^2)$ as described by (\ref{eq:X_density}).

\begin{figure}[htb]
    \centering
    \includegraphics[]{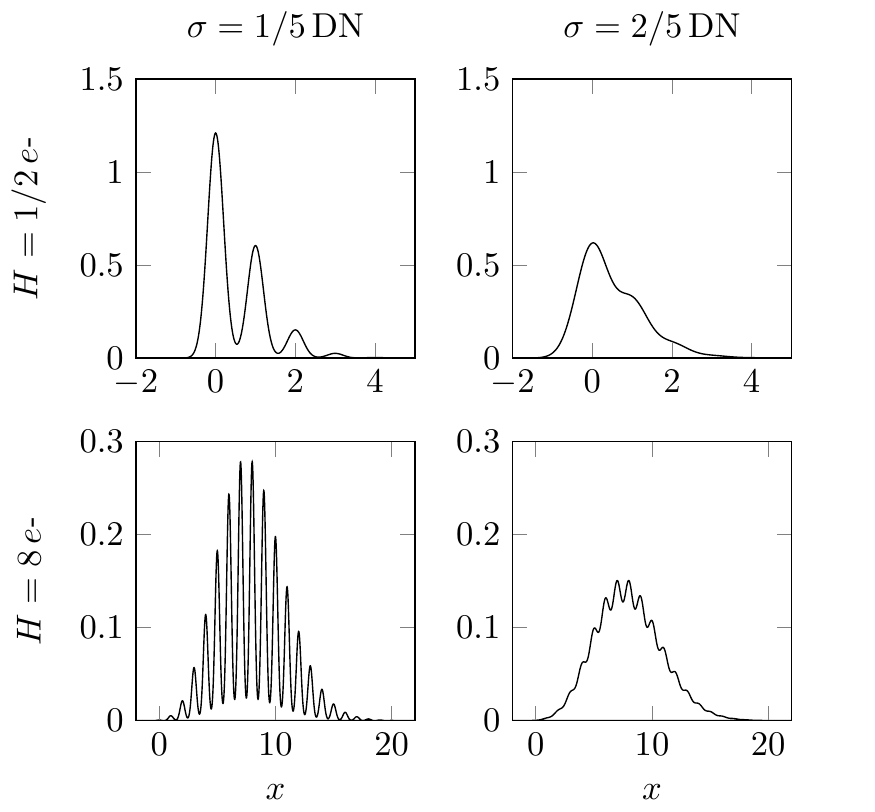}
    \caption{Plots of the PCD for various $H$ and $\sigma^2$ with $\mu=0$ and $g=1$ fixed.}
    \label{fig:PCD_plots}
\end{figure}

From (\ref{eq:X_density}) we see that the PCD is an infinite mixture of Gaussian components with $g$ controlling the spacing between each component, $\sigma$ as the width of each component, $\mu$ the location of the zeroth component, and $H$ the relative heights of the components. Figure \ref{fig:PCD_plots} plots the PCD for various $H$ and $\sigma^2$ with $\mu=0$ and $g=1$ fixed. From the figure we can see how the parameter $H$ controls the overall envelope of the distribution, while $\sigma$ determines whether the individual peaks may be resolved. As shown in the right column of Figure \ref{fig:PCD_plots}, as $\sigma$ increases, the contrast of the individual peaks is reduced. Typically, the term DSERN is assigned to pixels where $\sigma$ is small enough such that the peaks are clearly resolved \cite{fossum_2015}.


\section{The PCH-EM Algorithm}
\label{sec:PCH-EM_algorithm_derivation}

Looking back at (\ref{eq:X_density}), we can see that the PCD can be interpreted as the marginal density resulting from marginalizing $K$ out of the joint model $(X,K)$:
\begin{equation}
f_{XK}(x,k)=\frac{e^{-H}H^k}{k!}\phi(x;\mu+k/g,\sigma^2).    
\end{equation}
This reflects real-world data in the sense that we are only able to observe the pixel output $X$, while the number of free electrons $K$ is completely unknown. For this reason $K$ is a hidden (latent) variable in the PCD model. To see what impact hidden variables has on our approach for estimating $\theta$, Appendix \ref{sec:est_wo_hidden_variables} provides a derivation of the maximum likelihood estimator for $\theta$ in the hypothetical scenario where $K$ can be directly observed. As one can see in the appendix, the ability to directly observe $K$ rather unexpectedly results in a simple, closed-form estimator for $\theta$.

In the realistic scenario where $K$ is hidden, maximum likelihood estimation of $\theta$ becomes much less straightforward. To see why, we first denote $\mathbf x=(x_1,\dots,x_N)$, with $x_n\sim\operatorname{PCD}(H,g,\mu,\sigma^2)$, as a sample of $N$ observed values from a DSERN pixel. From this sample we may construct a likelihood function
\begin{equation}
    L(\theta|\mathbf x)\coloneqq f_X(\mathbf x|\theta)=\sum_\mathcal K f_{XK}(\mathbf x,\mathbf k|\theta),
\end{equation}
where we have used the shorthand notation $\mathbf k=(k_1,\dots,k_N)$, $\sum_\mathcal K=\sum_{k_1=0}^\infty\cdots\sum_{k_N=0}^\infty$, and $f_{XK}(\mathbf x,\mathbf k|\theta)=\prod_{n=1}^Nf_{XK}(x_n,k_n|\theta)$. Denoting $\ell(\theta|\mathbf x)=\log f_X(\mathbf x|\theta)$ as the log-likelihood function our goal is to solve the density estimation problem
\begin{equation}
    \tilde\theta=\argmax_\theta\ell(\theta|\mathbf x).
\end{equation}
Deriving the log-likelihood we find
\begin{equation}
    \ell(\theta|\mathbf x)=\sum_{n=1}^N\log\sum_{k=0}^\infty\frac{e^{-H}H^k}{k!}\phi(x_n;\mu+k/g,\sigma^2),
\end{equation}
which is problematic for deriving closed-form maximum likelihood estimators due to the series inside the logarithm. In situations like this, we can directly maximize $\ell(\theta|\mathbf x)$ through numerical methods such as gradient descent; however, this may be undesirable as one must calculate derivatives of the likelihood function and carefully control step size to ensure convergence.

A lesser-known, yet widely accepted, method for maximum likelihood estimation is that of the Expectation Maximization (EM) algorithm \cite{dempster_1977}. Instead of maximizing the log-likelihood function directly, the EM algorithm maximizes a related (often simpler) function to produce a sequence of estimators that converge to those of the maximum likelihood estimators. The key to the success of this method is that EM models the estimation problem as one containing hidden variables. Throughout the remainder of this section, we derive a custom-built EM algorithm for characterizing DSERN pixels, which takes into account the specific structure of the PCD in (\ref{eq:X_density}).


\subsection{Derivation of the PCH-EM Algorithm}

\subsubsection{Supporting Theory}

The key insight provided by the PCH-EM algorithm is that the log-likelihood can be written in an alternative form only made possible by knowing that the PCD model contains hidden variables. This alternative form then allows us to derive update equations that improve an initial estimate of $\theta$ in such a way as to guarantee an increase in the log-likelihood.

To derive this alternative form of the log-likelihood we first call on the definition of conditional density to write
\begin{equation}
    p_{K|X}(\mathbf k|\mathbf x,\theta)=\frac{f_{XK}(\mathbf x,\mathbf k|\theta)}{f_X(\mathbf x|\theta)},
\end{equation}
which upon taking the logarithm and rearranging gives
\begin{equation}
    \ell(\theta|\mathbf x)=\ell(\theta|\mathbf x,\mathbf k)-\log p_{K|X}(\mathbf k|\mathbf x,\theta),
\end{equation}
where $\ell(\theta|\mathbf x,\mathbf k)=\log f_{XK}(\mathbf x,\mathbf k|\theta)$. Multiplying both sides of this relation by $p_{K|X}(\mathbf k|\mathbf x,\theta^\prime)$ and summing over $\mathcal K$ we obtain the following alternative representation of the log-likelihood function that holds for any $\theta^\prime$ in the PCD parameter space \cite{Casella_MCbook}
\begin{equation}
    \label{eq:log_likelihood_alt_rep}
    \ell(\theta|\mathbf x)=\mathsf E_{\theta^\prime}(\ell(\theta|\mathbf x,\mathbf K))-\mathsf E_{\theta^\prime}(\log p_{K|X}(\mathbf K|\mathbf x,\theta)).
\end{equation}

Denoting the \emph{expected complete-data log-likelihood} as the first term in (\ref{eq:log_likelihood_alt_rep}), namely,
\begin{equation}
    \label{eq:expected_comp_log-likelihood}
    Q(\theta|\theta^{(t)})=\mathsf E_{\theta^{(t)}}(\ell(\theta|\mathbf x,\mathbf K)),
\end{equation}
the PCH-EM algorithm takes an initial estimate for the parameter $\theta^{(0)}=(H^{(0)},g^{(0)},\mu^{(0)},\sigma^{2(0)})$ and iterates between two steps: 
\begin{enumerate}
    \item The expectation (E) step to compute $Q(\theta|\theta^{(t)})$ and
    \item The Maximization (M) step, which maximizes $Q$ to update the estimate via
\begin{equation}
    \theta^{(t+1)}=\argmax_\theta Q(\theta|\theta^{(t)}).
\end{equation}
\end{enumerate}
Updating the parameter estimate in this way it is guaranteed in each iteration that \cite{casella_2002}
\begin{equation}
    \ell(\theta^{(t+1)}|\mathbf x)\geq\ell(\theta^{(t)}|\mathbf x)
\end{equation}
so that a local maxima of the log-likelihood is always achieved.  With the supporting theory we now present the details of the E- and M-step in the PCH-EM algorithm.


\subsubsection{E-Step}

The expectation step entails deriving $Q(\theta|\theta^{(t)})$. Substituting appropriate values into (\ref{eq:expected_comp_log-likelihood}) yields
\begin{equation}
    Q(\theta|\theta^{(t)})=\sum_\mathcal K p_{K|X}(\mathbf k|\mathbf x,\theta^{(t)})\log f_{XK}(\mathbf x,\mathbf k|\theta),
\end{equation}
which upon further expanding gives
\begin{multline}
    Q(\theta|\theta^{(t)})=\\
    \prod_{m=1}^N\sum_{k_m=0}^\infty p_{K|X}(k_m|x_m,\theta^{(t)})
    \sum_{n=1}^N\log f_{XK}(x_n,k_n|\theta).
\end{multline}
Bringing the sum w.r.t.~$n$ to the outside results in a lot of simplification. After interchanging the sums and substituting the expression for $f_{XK}(x_n,k|\theta)$ we obtain
\begin{equation}
    \label{eq:Q_fun}
    Q(\theta|\theta^{(t)})=
    \sum_{n=1}^N\sum_{k=0}^\infty \gamma_{nk}^{(t)}\log\left(\frac{e^{-H}H^k}{k!}\phi(x_n;\mu+k/g,\sigma^2)\right),
\end{equation}
where $\gamma_{nk}^{(t)}=p_{K|X}(k|x_n,\theta^{(t)})$ and
\begin{equation}
    \label{eq:p_K|X}
    p_{K|X}(k|x_n,\theta)=\frac{\frac{e^{-H}H^k}{k!}\phi(x_n;\mu+k/g,\sigma^2)}{\sum_{m=0}^\infty\frac{e^{-H}H^m}{m!}\phi(x_n;\mu+m/g,\sigma^2)}.
\end{equation}

Comparing the final expression in (\ref{eq:Q_fun}) to the complete log-likelihood of equation (\ref{eq:complete_data_log_likelihood}) in the appendix shows many similarities. The major difference between these expressions is that the $\mathds 1_{k_n=k}$ has been substituted for $\gamma_{nk}^{(t)}$. In the case where $K$ was known, the indicator function $\mathds 1_{k_n=k}$ can be interpreted as a degenerate probability distribution centered on the known value for each observed $k_n$. In comparison, in the PCH-EM algorithm we do not know $K$, so this degenerate distribution is replaced with $\gamma_{nk}^{(t)}$, representing the probability $K_n=k$ given the observed data $X_n=x_n$ and the current estimate of the parameter $\theta^{(t)}$. For this reason, the $\gamma_{nk}^{(t)}$ are commonly referred to as \emph{membership probabilities} as they assign the probability of $x_n$ belonging to each Gaussian component of the PCD.

\subsubsection{M-Step}

Maximization of $Q(\theta|\theta^{(t)})$ is a very similar process to maximizing the complete log-likelihood in (\ref{eq:complete_data_log_likelihood}). To maximize this function, we solve for the critical point $\nabla_\theta Q=0$. Solving this system of equations is fairly straightforward yet tedious. A proof can be found in Appendix \ref{sec:update_eqn_derivation}. Upon solving the system, we obtain the collection of update equations
\begin{subequations}\label{eq:update_eqns}
\begin{align}
H^{(t+1)} &=A^{(t)} \label{eq:H_update}\\
g^{(t+1)}&=\frac{B^{(t)}-H^{2(t+1)}}{C^{(t)}-\bar xH^{(t+1)}} \label{eq:g_update}\\
\mu^{(t+1)}&=\bar x-\frac{H^{(t+1)}}{g^{(t+1)}} \label{eq:mu_update}\\
\sigma^{2(t+1)} &= \frac{B^{(t)}}{g^{2(t+1)}}-2\frac{C^{(t)}}{g^{(t+1)}}+\overline{x^2}-\mu^{2(t+1)},\label{eq:sigma_update}
\end{align}
\end{subequations}
where $\bar x=\frac{1}{N}\sum_{n=1}^Nx_n$ and $\overline{x^2}=\frac{1}{N}\sum_{n=1}^Nx_n^2$ are the first two sample moments and
\begin{subequations}\label{eq:update_matrices}
\begin{align}
A^{(t)} &=\frac{1}{N}\sum_{n=1}^N\sum_{k=0}^\infty\gamma_{nk}^{(t)}k \label{eq:A_t}\\
B^{(t)} &=\frac{1}{N}\sum_{n=1}^N\sum_{k=0}^\infty\gamma_{nk}^{(t)}k^2 \label{eq:B_t}\\
C^{(t)} &=\frac{1}{N}\sum_{n=1}^Nx_n\sum_{k=0}^\infty\gamma_{nk}^{(t)}k. \label{eq:C_t}
\end{align}
\end{subequations}

Again, comparing these update equations to the closed-form maximum likelihood estimates in (\ref{eq:mle_solution}) shows many similarities as is expected. To check and see if these update equations make sense, notice that $A^{(t)}$, $B^{(t)}$, and $C^{(t)}$ have the form of Monte Carlo estimators for expected values w.r.t.~$X\sim\operatorname{PCD}(\theta)$, e.g.
\begin{equation}
    A^{(t)}\sim\mathsf E_\theta\left(\sum_{k=0}^\infty p_{K|X}(k|X,\theta^{(t)})k\right)
\end{equation}
as $N\to\infty$. Assuming we also have a good starting point so that $\theta^{(t)}\to\theta$ then produces the asymptotic approximations for large $N$ and large iteration number $t$
\begin{subequations}\label{eq:update_matrices_expectation_approx}
\begin{align}
A^{(t)} &\sim\mathsf EK &&\!\!\!\!\!\!\!\!\!\!\!\!\!\!\!\!\!\!\!\!=H \label{eq:A_t_approx}\\
B^{(t)} &\sim\mathsf EK^2 &&\!\!\!\!\!\!\!\!\!\!\!\!\!\!\!\!\!\!\!\!=H^2+H \label{eq:B_t_approx}\\
C^{(t)} &\sim\mathsf E(KX) &&\!\!\!\!\!\!\!\!\!\!\!\!\!\!\!\!\!\!\!\!=\frac{1}{g}(H^2+(1+\mu g)H). \label{eq:C_t_approx}
\end{align}
\end{subequations}
Likewise, the sample moments act as Monte Carlo estimators for the exact moments as $N\to\infty$ giving
\begin{subequations}\label{eq:update_matrices_expectation_approx2}
\begin{align}
\bar x &\sim\mathsf EX &&\!\!\!\!\!\!\!\!=\frac{1}{g}(H+\mu g) \label{eq:barx_approx}\\
\overline{x^2} &\sim\mathsf EX^2 &&\!\!\!\!\!\!\!\!=\frac{1}{g^2}(H^2+(1+2\mu g)H+(\mu g)^2+(\sigma g)^2). \label{eq:barx2_approx}
\end{align}
\end{subequations}
Substituting these asymptotic approximations into the right hand sides of (\ref{eq:update_eqns}) show that the updates are asymptotic to the parameters they estimate, e.g. $H^{(t+1)}\sim H$, $g^{(t+1)}\sim g$, and so on.

One additional benefit of the PCH-EM algorithm is that it also provides a means for estimating (demarginalizing) the hidden variable $K$ via the membership probabilities. Since $\gamma_{nk}^{(t)}$ represents a probability distribution for $K_n$ given $X_n=x_n$, we may estimate the hidden values $k_n$ via
\begin{equation}
    \label{eq:demarg_K_eqn}
    \tilde k_n=\argmax_k\gamma_{nk}^{(t)}
\end{equation}
and thus predict the number of free-electrons that generated each observation $x_n$.

\subsubsection{Final algorithm}

With the derivation of the E- and M-step now complete, the PCH-EM algorithm works by supplying a starting point $\theta^{(0)}=(H^{(0)},g^{(0)},\mu^{(0)},\sigma^{2(0)})$, then: 
\begin{enumerate}
    \item E: Compute $\gamma_{nk}^{(t)}$ by substituting $\theta^{(t)}$ into (\ref{eq:p_K|X}).
    \item M: Update $\theta^{(t)}\mapsto\theta^{(t+1)}$ with (\ref{eq:update_eqns}).
    \item Repeat until $Q(\theta^{(t+1)}|\theta^{(t)})-Q(\theta^{(t)}|\theta^{(t-1)})\leq\epsilon$.
\end{enumerate}


\section{Starting Point Estimation}
\label{sec:starting_points}

As is the case with numerical optimization methods, one of the challenges when working with the PCH-EM algorithm is the need for a starting point. While the PCH-EM update equations guarantee an increase in log-likelihood at each iteration, the effectiveness of the algorithm in achieving a global maximum, as compared to a local maximum, is sensitive to the initial starting point. Here we present a method for extracting the starting point directly from the original sample by relying on the properties of the PCD Fourier transform.


\subsection{Properties of the PCD Fourier Transform}

\begin{figure}[htb]
    \centering
    \includegraphics[]{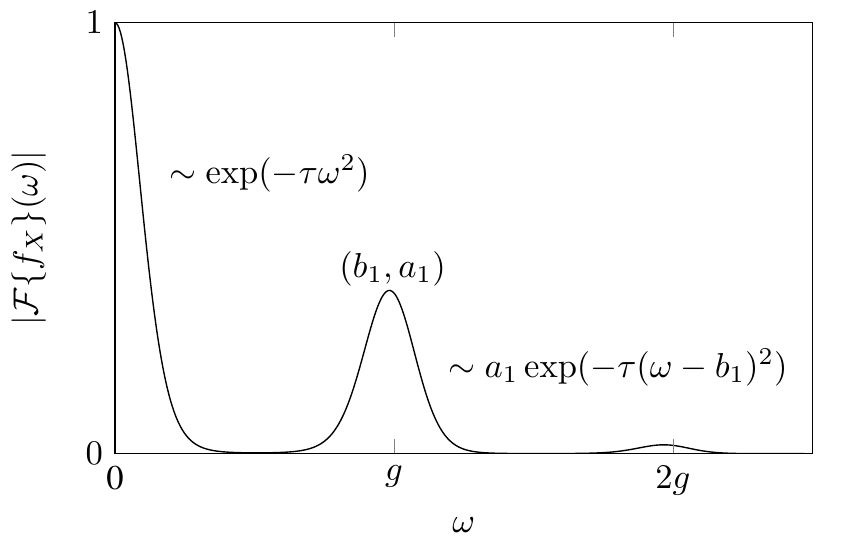}
    \caption{Graph of $|\mathcal F\{f_X\}(\omega)|$ versus $\omega$ showing the two most dominant peaks at $\omega=0$ and $\omega\approx b_1$ along with the approximate position of the secondary peak $(b_1,a_1)$.}
    \label{fig:FT_mag}
\end{figure}

We begin by considering the magnitude of the PCD Fourier transform $\mathcal F\{f_X\}(\omega)\coloneqq\mathsf E\exp(-2\pi i\omega X)$
\begin{equation}
    \label{eq:FT_magnitude}
    |\mathcal F\{f_X\}(\omega)|=\exp(H(\cos(2\pi\omega/g)-1)-2\pi^2\sigma^2\omega^2).
\end{equation}
The magnitude function in (\ref{eq:FT_magnitude}) is asymptotic to a Gaussian curve near integer multiples of $g$ in the sense that as $\omega\to ng$ $(n=0,1,2,\dots)$
\begin{equation}
    \label{eq:FT_mag_1st_peak}
    |\mathcal F\{f_X\}(\omega)|\sim a_n\exp(-\tau(\omega-b_n)^2),
\end{equation}
with $\tau=2 \pi^2 (H/g^2+\sigma^2)$,
\begin{equation}
    a_n=\exp\left(-2\pi^2\left(Hn^2-\frac{(Hn)^2}{g^2(H/g^2+\sigma^2)}\right)\right),   
\end{equation}
and
\begin{equation}
    b_n=\frac{Hn}{g(H/g^2+\sigma^2)}.    
\end{equation}
From this observation we expect to find local maxima (peaks) in the magnitude function at $\omega\approx b_n$. Figure \ref{fig:FT_mag} depicts a graph of $|\mathcal F\{f_X\}(\omega)|$ showing the two most dominant peaks at $\omega=0$ and $\omega\approx b_1$ along with the approximate position of the secondary peak $(\omega,|\mathcal F\{f_X\}(\omega)|)=(b_1,a_1)$.

The existence of the secondary peak at $\omega\approx b_1$ depends on the values of quanta exposure and read noise. To see why, we evaluate $\partial_\omega|\mathcal F\{f_X\}(\omega)|=0$, which given $\omega>0$ simplifies to
\begin{equation}
    \operatorname{sinc}(2\pi\omega/g)+\frac{(\sigma g)^2}{H}=0,
\end{equation}
where $\operatorname{sinc}x=\sin x/x$. In order for a secondary peak to exist, this equation must have at least two solutions on $\omega>0$, which occurs only when
\begin{equation}
    \frac{(\sigma g)^2}{H}<|\min_{x>0}\operatorname{sinc}x|=0.217\dots
\end{equation}
In other words, the magnitude function at $\omega\approx b_1$ is only a peak if $H$ is about five times larger than $(\sigma g)^2$. When this is not the case, we cannot reliably extract the starting point from the observed data. From the properties outlined here we can extract the starting point as follows.


\subsection{Staring Point Extraction Procedure}

Given a set of $N$ observations $\mathbf x=(x_1,\dots,x_N)$ with $x_n\sim\operatorname{PCD}(H,g,\mu,\sigma^2)$, we begin the process of extracting the starting point $\theta^{(0)}$ by creating a density normalized PCH via
\begin{equation}
    \tilde f_X(n)=\frac{1}{N}\sum_{k=1}^N\mathds 1_{x_k=n},
\end{equation}
where $n\in\{\min(\mathbf x),\min(\mathbf x)+1,\dots,\max(\mathbf x)\}$. The number of bins in the density normalized PCH is denoted $N_b=\max(\mathbf x)-\min(\mathbf x)+1$. Next, we evaluate the Discrete Fourier transform (DFT)
\begin{equation}
    \mathcal F\{\tilde f_X\}(\omega_n)=\sum_{k}\tilde f_X(k)\exp(-2\pi i (k-1)\omega_n),
\end{equation}
where $\omega_n=(n-1)/N_b$.

The location of the secondary peak in the DFT magnitude $|\mathcal F\{\tilde f_X\}(\omega_n)|$ yields a single point, which encodes estimates of $a_1$ and $b_1$, namely, $(\omega_\text{peak},|\mathcal F\{\tilde f_X\}(\omega_\text{peak})|)=(\tilde b_1,\tilde a_1)$. Likewise, with the help of (\ref{eq:FT_mag_1st_peak}) an estimate of $\tau$ can be extracted by fitting the quadratic function $-\tau\omega^2$ to the logarithm of points in $|\mathcal F\{\tilde f_X\}(\omega_n)|$ belonging to the $\omega=0$ peak.  Equating $\tilde a_1$, $\tilde b_1$, and $\tilde\tau$ with their exact expressions yields a system of equations that can be inverted to obtain initial estimates of $H$, $g$, and $\sigma^2$:
\begin{subequations}\label{eq:update_matrices}
\begin{align}
\tilde H &=\frac{\tilde b_1^2\tilde\tau-\log\tilde a_1}{2\pi^2} \label{eq:initial_H}\\
\tilde g &=2\pi^2\frac{\tilde H}{\tilde b_1\tilde\tau} \label{eq:initial_g}\\
\tilde\sigma^2 &=\frac{\tilde\tau}{2\pi^2}-\frac{\tilde H}{\tilde g^2}. \label{eq:initial_v}
\end{align}
\end{subequations}
The final starting points $H^{(0)}$, $g^{(0)}$, and $\sigma^{2(0)}$ are then computed by fitting the full model (\ref{eq:FT_magnitude}) to $|\mathcal F\{\tilde f_X\}(\omega_n)|$ using nonlinear least squares with $\tilde H$, $\tilde g$, and $\tilde\sigma^2$ as starting values.

As for the starting value of $\mu$, we obtain an initial estimate via
\begin{equation}
    \tilde\mu=\bar x-\frac{H^{(0)}}{g^{(0)}}.
\end{equation}
If the estimate $\tilde\mu$ differs from the exact value of $\mu$ by approximately $1/(2g)$, e.g.~$\tilde\mu\approx\mu\pm 1/(2g)$, supplying this estimate to the PCH-EM algorithm will result in slow convergence. This can be mostly circumvented by first constructing the autocorrelation function
\begin{equation}
    R(t)=\sum_k\tilde f_X(k)f_X(k-t|\tilde\mu,H^{(0)},g^{(0)},\sigma^{2(0)})
\end{equation}
and then extracting a correction factor
\begin{equation}
    \text{correction}=\argmax_t R(t)
\end{equation}
to refine the initial estimate via
\begin{equation}
    \mu^{(0)}=\tilde\mu+\text{correction}.
\end{equation}


\section{Implementation and Examples}
\label{sec:implementation_and_examples}

To demonstrate the utility of the derived results, the starting point algorithm as well as the PCH-EM algorithm were implemented in MATLAB.  This code can be freely downloaded from the MathWorks file exchange \cite{PCHEM_code}. Using this code we simulated the TPG jot presented in \cite{starkey_2016} with the parameters in Table \ref{tab:sim_parameters}. A sample of $N=1000$ observations for the simulated jot were generated according to (\ref{eq:PCD_RV_definition}), namely,
$X=\lceil(K+R)/g\rfloor$, where $K\sim\operatorname{Poisson}(H)$ and $R\sim\mathcal N(\mu_R,\sigma_R)$.

\begin{table}[htb]
\begin{center}
\caption{Simulation parameters.}
\label{tab:sim_parameters}
\begin{tabular}{| c | c | c |}
\hline
Quantity & Symbol & Value\\
\hline
\hline
sample size & $N$ & $1000\, (-)$\\
\hline
dark current & $i_d$ & $0.12\,(e\text{-}/s)$\\
\hline
integration time & $t$ & $0.1\,(s)$\\
\hline
dark quanta exposure & $H_d=i_d\times t$ & $0.012\,(e\text{-})$\\
\hline
photon quanta exposure & $H_\gamma$ & $6.858\,(e\text{-})$\\
\hline
total quanta exposure & $H=H_\gamma+H_d$ & $6.87\,(e\text{-})$\\
\hline
bias & $\mu_R$ & $1.82\,(e\text{-})$\\
\hline
read noise & $\sigma_R$ & $0.26\,(e\text{-})$\\
\hline
conversion gain & $g$ & $0.0433\,(e\text{-}/\mathrm{DN})$\\
\hline
quantization noise & $\sigma_Q$ & $0.2887\,(\mathrm{DN})$\\
\hline
bias (sensor units) & $\mu=\mu_R/g$ & $42\,(\mathrm{DN})$\\
\hline
read noise (sensor units) & $\sigma=((\sigma_R/g)^2+\sigma_Q^2)^{1/2}$ & $6.0069\,(\mathrm{DN})$\\
\hline
\end{tabular}
\end{center}
\end{table}

\begin{figure}[htb]
    \centering
    \includegraphics[]{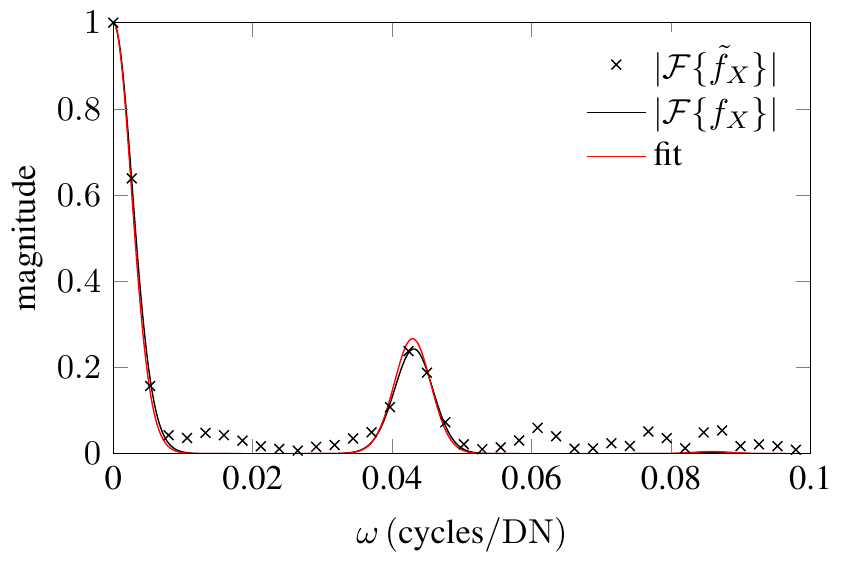}
    \caption{Magnitude of PCH DFT $(|\mathcal F\{\tilde f_X\}|)$ compared against the exact PCD Fourier transform magnitude $(|\mathcal F\{f_X\}|)$ and fitted magnitude.}
    \label{fig:plot_StartingPoints_results}
\end{figure}

The first step in the process is to generate the starting point. The algorithm for generating the starting point requires no user input and only accepts the raw sensor data.  Figure \ref{fig:plot_StartingPoints_results} presents the result of the starting point algorithm showing the simulated PCH DFT magnitude compared to the magnitude of the exact PCD Fourier transform and the estimated fit using our starting point algorithm.

\begin{figure}[htb]
    \centering
    \includegraphics[]{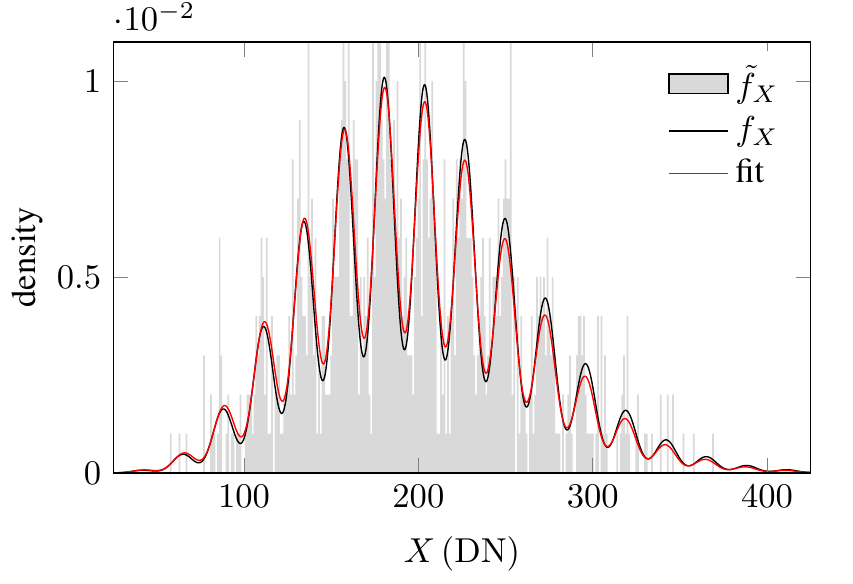}
    \caption{Simulated PCH data $(\tilde f_X)$ versus the exact PCD $(f_X)$ and fitted PCD generated by the PCH-EM algorithm.}
    \label{fig:plot_PCHEM_results}
\end{figure}

The estimated starting point, along with the raw data $\mathbf x=(x_1,\dots,x_{1000})$, were then fed into the PCH-EM algoithm. The algorithm converged in just two iterations, which took approximately 0.042 seconds to complete running on a Intel Core i7 processor. Figure \ref{fig:plot_PCHEM_results} plots the simulated PCH data along with the exact PCD (according to Table \ref{tab:sim_parameters}) and fitted PCD generated by the PCH-EM algorithm parameter estimate. Table \ref{tab:estimated_parameters} also shows the estimated parameters along with their exact values and percent error. From these results we can see that we were successfully able to estimate all four parameters from just $1000$ observations.  In particular, we were able to estimate the conversion gain with less than $1\%$ error. From these estimates we were also able to estimate the read noise $\sigma g$ yielding the value $6.2585\,e\text{-}$, which results in $4.67\%$ error. For comparison, we generated a second dark sample of $M=1000$ observations according to the PCD model
$Y=\lceil (K+R)/g\rfloor$, where $K\sim\operatorname{Poisson}(H_d)$ and $R\sim\mathcal N(\mu_R,\sigma_R^2)$. We then calculated the conversion gain using the two-sample PT gain estimator described in \cite{hendrickson_22}
\begin{equation}
    \tilde g=\frac{\bar x-\bar y}{\hat x-\hat y},
\end{equation}
where $\bar x$ is the sample mean of the illuminated data, $\hat x$ is the sample variance of the illuminated data, and likewise for the dark data.  This estimator gave an estimate of of the conversion gain equal to $\tilde g=0.0489$ resulting in a substantially larger error of $12.8\%$.

\begin{table}[htb]
\begin{center}
\caption{Estimated parameters for simulated jot.}
\label{tab:estimated_parameters}
\begin{tabular}{| c | c | c | c |}
\hline
Quantity & Estimate & Exact & Error\\
\hline
\hline
$H\,(e\text{-})$ & $6.7394$ & $6.87$ &$1.90\%$\\
\hline
$g\,(e\text{-}/\mathrm{DN})$ & $0.0435$ & $0.0433$ &$0.46\%$\\
\hline
$\mu\,(\mathrm{DN})$ & $42.869$ & $42$ &$2.07\%$\\
\hline
$\sigma\,(\mathrm{DN})$ & $6.2585$ & $6.0069$ &$4.19\%$\\
\hline
\end{tabular}
\end{center}
\end{table}

\begin{figure}[htb]
    \centering
    \includegraphics[]{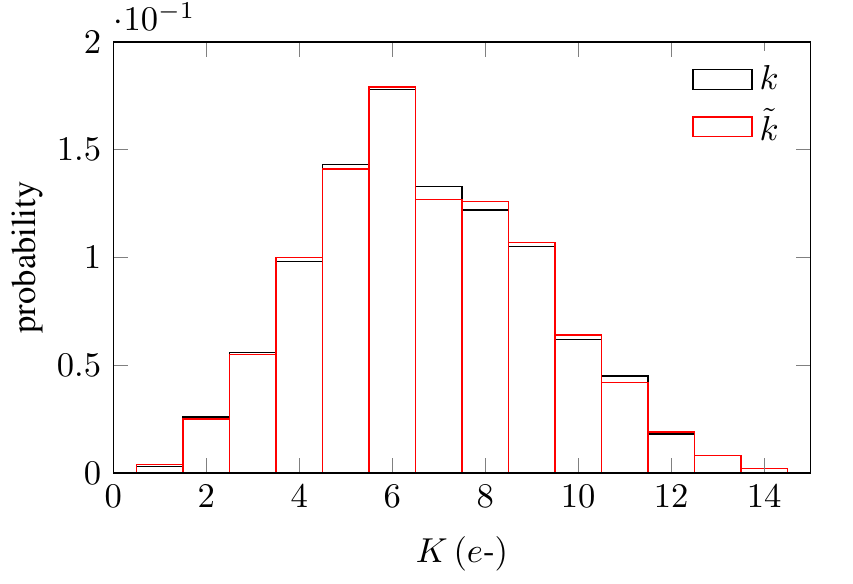}
    \caption{Histogram of estimates for the hidden variable $K$ compared to the histogram of their exact values.}
    \label{fig:plot_demarg_K}
\end{figure}

After the PCH-EM algorithm was complete, we took the membership probabilities, $\gamma_{nk}^{(t)}$, from the final iteration and used them to estimate the values hidden variable $K$ corresponding to each $x_n$ via (c.f.~(\ref{eq:demarg_K_eqn}))
\begin{equation}
    \tilde k_n=\argmax_k\gamma_{nk}^{(t)}.
\end{equation}
These estimates are plotted as a histogram in Figure \ref{fig:plot_demarg_K} along with the exact values of $k_n$, which were hidden from the algorithm.  As we observe from the figure, the PCH-EM algorithm was able to demarginalize the hidden variable $K$ as indicated by the fact that the histogram of $\tilde k_n$ is in close agreement with the histogram of the exact values.  For this dataset, $93.7\%$ of the estimates $\tilde k_n$ agreed with the exact values $k_n$.

To evaluate the algorithm's behavior for the chosen parameters, we repeated this experiment $10,000$ times at $N=1000$ and again at $N=5000$, observing the distribution of starting and final estimates. Tables \ref{tab:MC_result_N1000}-\ref{tab:MC_result_N5000} compare the exact parameter values to the estimated starting point and final parameter estimates ($\pm$ one standard deviation) for $N=1000$ and $N=5000$, respectively. From Table \ref{tab:MC_result_N1000}, we see all parameters are approximately unbiased; however, we note a couple of anomalies, in particular, high variance in the starting values and final estimates of $\mu$ and increased variance in the final estimates of $H$ compared to the starting values. These anomalies disappear in the corresponding data for $N=5000$.

\begin{figure}[htb]
    \centering
    \includegraphics[]{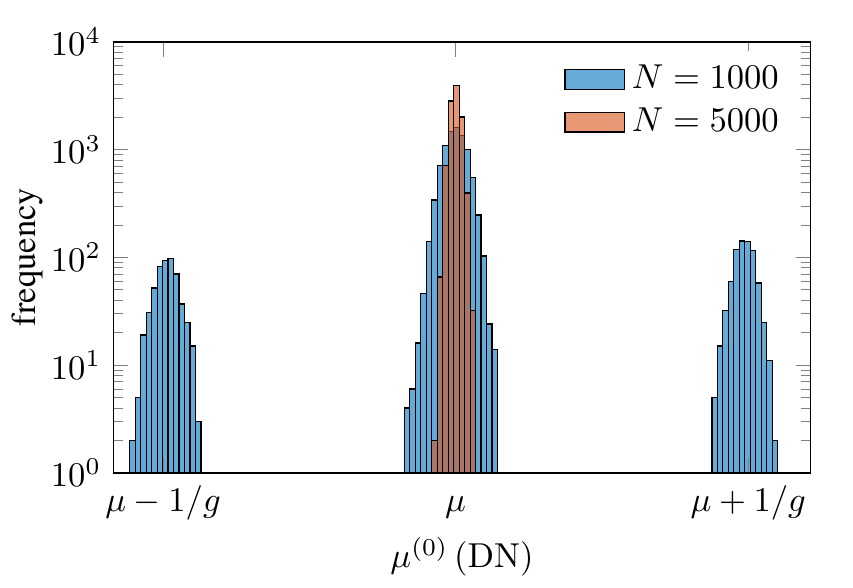}
    \caption{Histograms of $\mu^{(0)}$ for $10,000$ Monte Carlo experiments using $N=1000$ and $N=5000$.}
    \label{fig:plot_mu_start}
\end{figure}

The source of the anomalies for the $N=1000$ data comes from the starting values for $\mu$. Figure \ref{fig:plot_mu_start} plots histograms of the $10,000$ starting values $\mu^{(0)}$ for the $N=1000$ and $N=5000$ runs. We can see that at $N=1000$ the starting values are trimodal: one mode centered about the correct value of $\mu$ and the other two modes centered approximately around $\mu\pm1/g$. We note the use of a logarithmic scale to enhance the appearence of these additional modes. When the starting values for $\mu$ were in one of these secondary modes the final estimates for $H$ would also be negatively affected. However, we note that the estimates for $g$ and $\sigma^2$ are quite robust and yield satisfactory results even when a starting point for $\mu$ ends up in one of these incorrect modes. These problems are much less likely when increasing the sample size to $N=5000$.  As one additional exercise, we also repeated the experiment another $10,000$ times for $N=1000$ but this time we replaced the starting value for $\mu$ by the sample mean of $M=1000$ dark observations $Y\sim\operatorname{PCD}(H_d,g,\mu,\sigma^2)$. When we did this the multi-modal behavior completely disappeared from both the starting values and final estimates. This suggests that when working with small $N$, the algorithm is dramatically improved by estimating the starting value for $\mu$ from an independent dark sample.

\begin{table}[htb]
\begin{center}
\caption{Comparison of starting values and final parameter estimates to exact value for $N=1000$.}
\label{tab:MC_result_N1000}
\begin{tabular}{| c | c | c | c |}
\hline
Quantity & Exact & Start & Final\\
\hline
\hline
$H\,(e\text{-})$ & $6.87$ & $6.8419\pm 0.3380$ &$6.8515\pm 0.3629$\\
\hline
$g\,(e\text{-}/\mathrm{DN})$ & $0.0433$ & $0.0433\pm 0.0002$ &$0.0433\pm 0.0002$\\
\hline
$\mu\,(\mathrm{DN})$ & $42$ & $42.4309\pm 8.1255$ &$42.4042\pm 8.0173$\\
\hline
$\sigma^2\,(\mathrm{DN}^2)$ & $36.083$ & $35.8996\pm 2.1886$ &$36.0569\pm 2.1298$\\
\hline
\end{tabular}
\end{center}
\end{table}

\begin{table}[htb]
\begin{center}
\caption{Comparison of starting values and final parameter estimates to exact value for $N=5000$.}
\label{tab:MC_result_N5000}
\begin{tabular}{| c | c | c | c |}
\hline
Quantity & Exact & Start & Final\\
\hline
\hline
$H\,(e\text{-})$ & $6.87$ & $6.8622\pm 0.1545$ &$6.8700\pm 0.0461$\\
\hline
$g\,(e\text{-}/\mathrm{DN})$ & $0.0433$ & $0.0433\pm 0.0001$ &$0.0433\pm 0.0001$\\
\hline
$\mu\,(\mathrm{DN})$ & $42$ & $42.0050\pm 0.7301$ &$42.0073\pm 0.6998$\\
\hline
$\sigma^2\,(\mathrm{DN}^2)$ & $36.083$ & $36.0539\pm 0.9936$ &$36.0687\pm 0.9389$\\
\hline
\end{tabular}
\end{center}
\end{table}

\section{Conclusion}

In this work we have developed the PCH-EM algorithm for estimating key performance parameters of DSERN pixels. A model for DSERN sensor data was derived in the form of the PCD, which was in turn was used to derive the PCH-EM algorithm from the principle of expectation maximization. A method for estimating the starting point for the PCH-EM algorithm was also discussed. These algorithms were implemented in MATLAB and Monte Carlo experiments validated the effectiveness of the algorithms in estimating key performance parameters of DSERN pixels. The specific parameters selected to demonstrate the methods came from \cite{starkey_2016}, however we encourage the interested reader to download the provided source code and adjust the parameters to their specific device \cite{PCHEM_code}. A useful feature of using the provided Monte Carlo method is the ability to conduct sensitivity analysis on potential experimental parameters, e.g.~sample sizes.

Future work on the PCH-EM algorithm can be further extended to include two-samples: one captured under illumination and another under dark conditions. Doing so would allow us to separate out the effects of photon interactions and dark current by estimating $H_\gamma$ and $H_d$ separately, which may be of interest to the community. We also note from our simulation results that the introduction of a dark sample stabilized initial estimates of $\mu$ so we might expect a two-sample version of PCH-EM to be more stable compared to the current one-sample method when working with small sample sizes. Additionally, it is important to investigate the performance of the PCH-EM algorithm at higher read noise levels where peaks in the PCD are no longer discernible, identifying areas where different methods are valid.


\section*{Acknowledgments}
The authors would like thank Nico Schl\"{o}mer for his {\ttfamily{matlab2tikz}} function, which was used to create the figures throughout this work \cite{schlomer_2021}.


\appendices

\section{Parameter Estimation Without Hidden Variables}
\label{sec:est_wo_hidden_variables}

Suppose $K$ is not hidden so that we could directly observe the complete data $(\mathbf x,\mathbf k)=((x_1,k_1),\dots,(x_N,k_N))$. With this sample, a maximum likelihood estimator for the parameter $\theta$ is very easy to obtain. The likelihood function takes the form
\begin{equation}
    L(\theta|\mathbf x,\mathbf k)=\prod_{n=1}^N\prod_{k=0}^\infty \left(\frac{e^{-H}H^k}{k!}\phi(x_n;\mu+k/g,\sigma^2)\right)^{\mathds 1_{k_n=k}},
\end{equation}
which upon taking the logarithm yields the corresponding log-likelihood function
\begin{equation}
    \label{eq:complete_data_log_likelihood}
    \ell(\theta|\mathbf x,\mathbf k)=\sum_{n=1}^N\sum_{k=0}^\infty\mathds 1_{k_n=k}\log\left(\frac{e^{-H}H^k}{k!}\phi(x_n;\mu+k/g,\sigma^2)\right).
\end{equation}
Here, $\mathds 1_A$ denotes the indicator function which is equal to one when $A$ is true and zero otherwise. By the definition of the indicator function, the log-likelihood then simplifies to
\begin{equation}
    \ell(\theta|\mathbf x,\mathbf k)=\sum_{n=1}^N\log\left(\frac{e^{-H}H^{k_n}}{k_n!}\phi(x_n;\mu+k_n/g,\sigma^2)\right).
\end{equation}
The maximum likelihood estimate of $\theta$ then comes from solving for the critical point $\nabla_\theta\ell=0$. Equating the appropriate derivatives to zero and solving the resulting system of equations we obtain
\begin{subequations}\label{eq:mle_solution}
\begin{align}
\tilde H &=A \label{eq:H_est}\\
\tilde g &=\frac{B-\tilde H^2}{C-\bar x\tilde H} \label{eq:g_est}\\
\tilde\mu &=\bar x-\frac{\tilde H}{\tilde g} \label{eq:mu_est}\\
\tilde\sigma^2 &= \frac{B}{\tilde g^2}-2\frac{C}{\tilde g}+\overline{x^2}-\tilde\mu^2,\label{eq:sigma_est}
\end{align}
\end{subequations}
where
\begin{subequations}\label{eq:update_matrices}
\begin{align}
A &=\frac{1}{N}\sum_{n=1}^Nk_n \label{eq:A}\\
B &=\frac{1}{N}\sum_{n=1}^Nk_n^2 \label{eq:B}\\
C &=\frac{1}{N}\sum_{n=1}^Nx_nk_n. \label{eq:C}
\end{align}
\end{subequations}

So in the case where $K$ is not hidden (it can be directly observed), closed-form maximum likelihood estimators for the PCD parameter are nearly trivial to derive.


\section{Derivation of PCH-EM Update Equations}
\label{sec:update_eqn_derivation}

We begin with the expression for $Q(\theta|\theta^{(t)})$ in (\ref{eq:Q_fun}) to write
\begin{multline}
    Q(\theta|\theta^{(t)})=
    \sum_{n=1}^N\sum_{k=0}^\infty \gamma_{nk}^{(t)}\biggl(-H+k\log H\\
    -\frac{1}{2}\log\sigma^2-\frac{(x_n-\mu-k/g)^2}{2\sigma^2}+C\biggr),
\end{multline}
where $C$ is a constant independent of $\theta$. The update equations are then derived by solving the system of equations $\nabla_\theta Q=0$. Taking the derivative of $Q$ w.r.t.~$H$, equating with zero, and simplifying yields
\begin{equation}
    \sum_{n=1}^N\sum_{k=0}^\infty \gamma_{nk}^{(t)}(H-k)=0.
\end{equation}
Because the $\gamma_{nk}^{(t)}$ represent probabilities w.r.t.~the index $k$ we have $\sum_{k=0}^\infty \gamma_{nk}^{(t)}=1$ so that $\sum_{n=1}^N\sum_{k=0}^\infty \gamma_{nk}^{(t)}=N$. Recalling the definition of $A^{(t)}$ then leads to the solution 
\begin{equation}
    H^{(t+1)}=A^{(t)}.
\end{equation}

Next we evaluate $\partial Q/\partial\mu=0$, which after some simplification gives
\begin{equation}
    \label{eq:mu_update_equation_unsolved}
    \sum_{n=1}^N\sum_{k=0}^\infty \gamma_{nk}^{(t)}(x_n-\mu-k/g)=0.
\end{equation}
Expanding and simplifying we obtain an expression for $\mu$ in terms of $g$, namely,
\begin{equation}
    \mu^{(t+1)}=\bar x-\frac{H^{(t+1)}}{g^{(t+1)}}.
\end{equation}

To find the update equation for $g$ we repeat the process by evaluating and simplifying $\partial Q/\partial g=0$ to find
\begin{equation}
    \sum_{n=1}^N\sum_{k=0}^\infty \gamma_{nk}^{(t)}(x_n-\mu-k/g)k=0.
\end{equation}
Substituting $\mu=\bar x-H^{(t+1)}/g$ then gives us an equation with one unknown (unknown in $g$). Using the definitions of $A^{(t)}$ (which equals $H^{(t+1)}$), $B^{(t)}$, and $C^{(t)}$ we are able to write
\begin{equation}
    C^{(t)}-\bar x H^{(t+1)}+\frac{1}{g} H^{2(t+1)}-\frac{1}{g} B^{(t)}=0.
\end{equation}
This equation is then easily solved for $g$ yielding the update equation $g^{(t+1)}$ and subsequently $\mu^{(t+1)}$.

Lastly we evaluate $\partial Q/\partial\sigma^2=0$ and simplify to obtain
\begin{equation}
    \sum_{n=1}^N\sum_{k=0}^\infty \gamma_{nk}^{(t)}(\sigma^2-(x_n-\mu-k/g)^2)=0.
\end{equation}
Solving this equation for $\sigma^2$ gives
\begin{equation}
    \sigma^{2(t+1)}=\frac{1}{N}\sum_{n=1}^N\sum_{k=0}^\infty \gamma_{nk}^{(t)}(x_n-\mu^{(t+1)}-k/g^{(t+1)})^2.
\end{equation}
Expanding the trinomial term and simplifying we find after much algebra
\begin{multline}
    \sigma^{2(t+1)}=\frac{B^{(t)}}{g^{2(t+1)}}-2\frac{C^{(t)}}{g^{(t+1)}}+\overline{x^2}+\mu^{2(t+1)}\\
    -2\mu^{(t+1)}\frac{1}{N}\sum_{n=1}^N\sum_{k=0}^\infty \gamma_{nk}^{(t)}(x_n-k/g^{(t+1)}).
\end{multline}
Identifying the remaining double sum as $\mu^{(t+1)}$ (c.f.~(\ref{eq:mu_update_equation_unsolved})) then leads to the final solution for $\sigma^{2(t+1)}$.


\bibliographystyle{IEEEtran}
\bibliography{sources}



\begin{IEEEbiography}[{\includegraphics[width=1in,height=1.25in,clip,keepaspectratio]{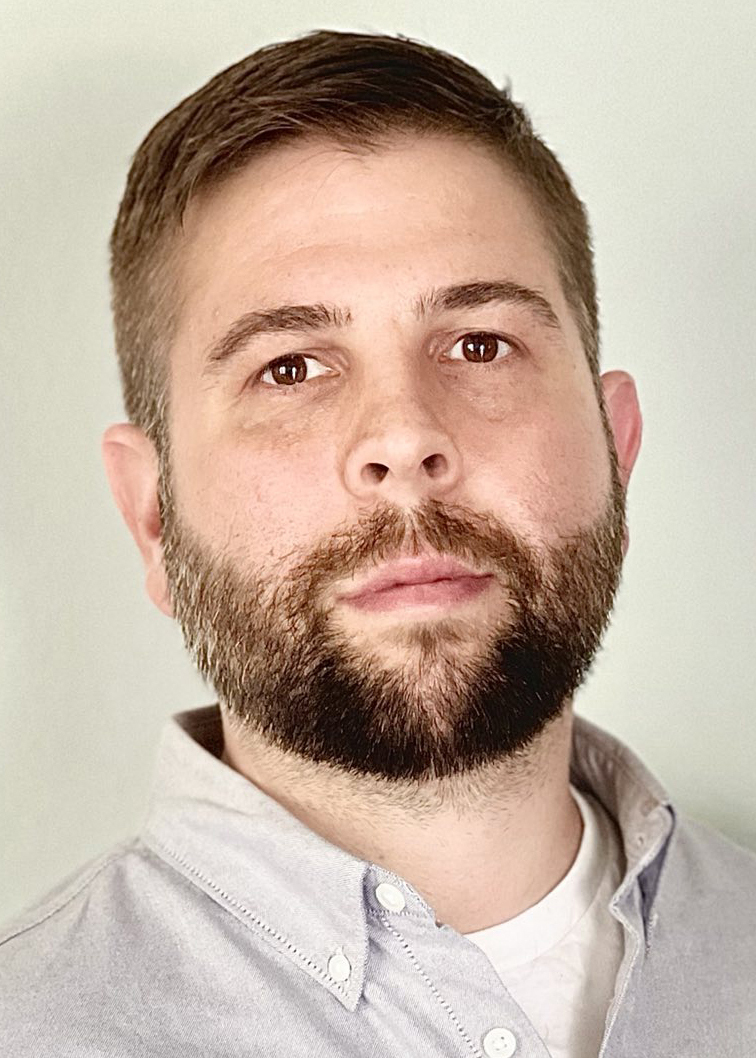}}]{Aaron Hendrickson}
received the B.S.~degree in Imaging and Photographic Technology from the Rochester Institute of Technology, Rochester, NY, USA, in 2011, and the M.S.~degree in Applied and Computational Mathematics from Johns Hopkins University, Baltimore, MD, USA, in 2020. He is currently a research mathematician working for the U.S.~Department of Defense at NAWCAD's DAiTA group. His research focus is in developing theoretical foundations for image sensor characterization methods.
\end{IEEEbiography}

\begin{IEEEbiography}[{\includegraphics[width=1in,height=1.25in,clip,keepaspectratio]{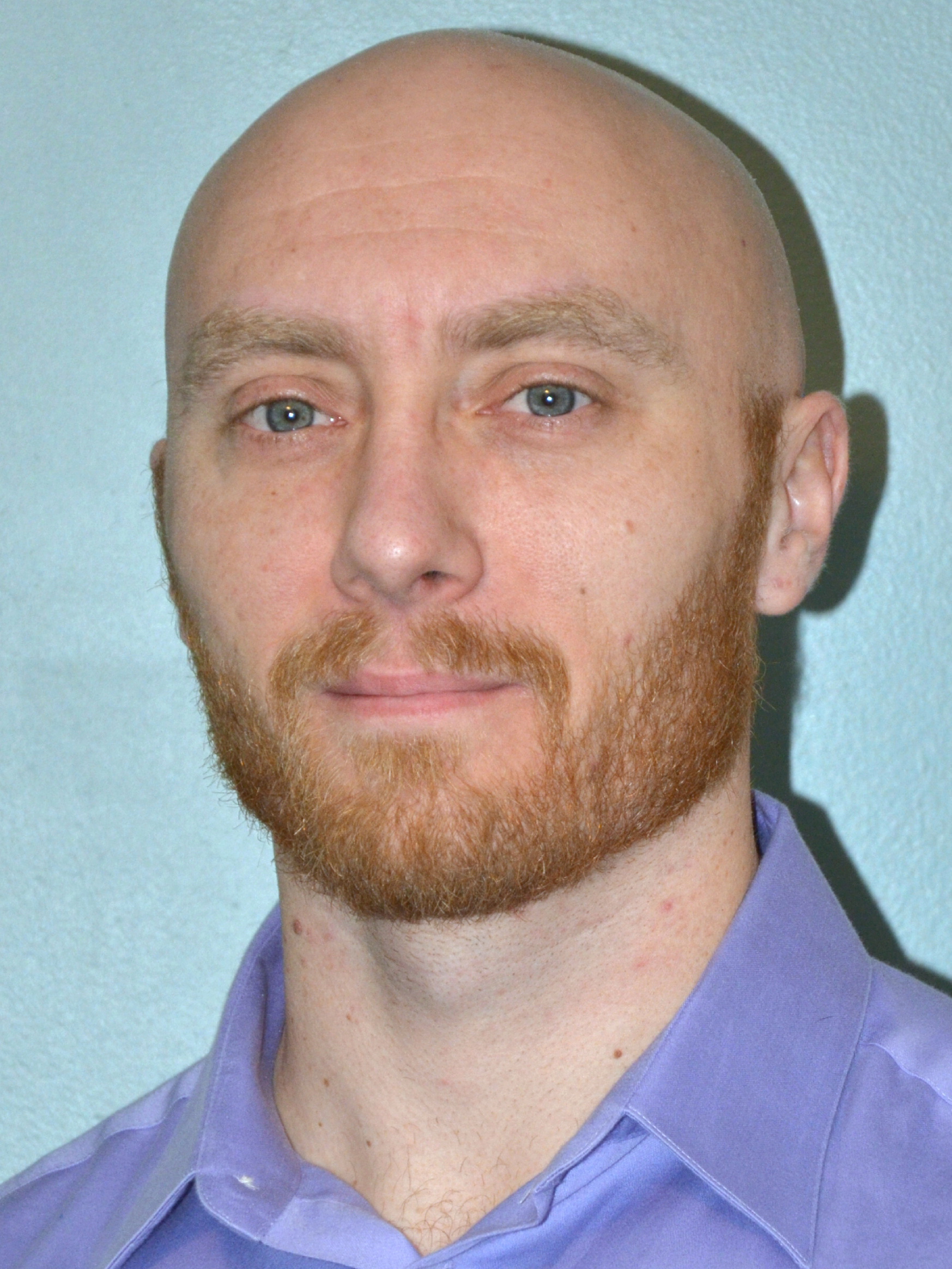}}]{David P. Haefner}
received his B.S.~in Physics from ETSU in 2004, a Ph.D.~in Optics from the UCF’s CREOL in 2010, a M.S.~in Electrical Engineering, and a M.S.~in Mechanical Engineering from CUA in 2014 and 2015, respectively. Since 2010 he has worked at the U.S. Army C5ISR Center. His current research spans electro-optic imaging system measurement for performance predictions and new measurement development.
\end{IEEEbiography}

\vfill

\end{document}